\documentclass[runningheads]{llncs}
\usepackage{graphicx}
\usepackage{amsmath,amssymb,mathrsfs,framed,esint,slashed}
\usepackage[colorlinks]{hyperref}
\usepackage{color}
\usepackage[all]{xy}

\newcommand{\de}{{\rm d}}

\newcommand{\beq}{\begin{equation}}
\newcommand{\eeq}{\end{equation}}

\begin{document}
\title{From quantum hydrodynamics\\to Koopman wavefunctions II\thanks{CT acknowledges partial support by the Royal Society and the Institute of Mathematics and its Applications, UK.}
}
%
%
\author{
\makebox{Cesare Tronci\inst{1}
\and Fran\c{c}ois Gay-Balmaz\inst{2}
}}%
\authorrunning{F. Gay-Balmaz and C. Tronci}
%
\institute{$^1$Department of Mathematics, University of Surrey, Guildford, UK\\
\makebox{Department of Physics and Engineering Physics, Tulane University, New Orleans, USA}\\
 \email{c.tronci@surrey.ac.uk}\\
 \medskip
$^2$CNRS \& \'Ecole Normale Sup\'erieure, Paris, France \\
\email{francois.gay-balmaz@lmd.ens.fr}}
\maketitle              
\begin{abstract} Based on the Koopman-van Hove (KvH) formulation of classical mechanics introduced in Part I, we formulate a Hamiltonian model for hybrid quantum-classical systems. This is obtained by writing the KvH wave equation for two classical particles and applying  canonical quantization to one of them. We illustrate several geometric properties of the model regarding the associated  quantum,  classical, and hybrid densities. After presenting the quantum-classical Madelung transform, the joint quantum-classical distribution is shown to arise as a momentum map for a unitary action naturally induced from the van Hove representation on the hybrid Hilbert space. While the quantum density matrix is positive by construction, no such result is currently available for the classical density. However, here we present a class of hybrid Hamiltonians whose flow preserves the sign of the classical density. Finally, we provide a simple closure model based on momentum map structures.

\keywords{Mixed quantum-classical  dynamics \and Koopman wavefunctions  \and Hamiltonian systems \and momentum maps.}
\end{abstract}

\section{Introduction}
Following the first part of this work, here we deal with the dynamics of hybrid systems where one subsystem is classical while the other is quantum. While concrete occurrences of quantum-classical physical systems still need to be identified, quantum-classical models have been sought for decades in the field of quantum chemistry. In this context,  nuclei are treated as classical while electrons retain their fully quantum nature. A celebrated example is given by  Born-Oppenheimer molecular dynamics.
The formulation of hybrid classical-quantum dynamics is usually based on fully quantum treatments, in which
some kind of factorization ansatz is invoked on the wavefunction. This ansatz is then followed by a classical
limit on the factor that is meant to model the classical particle.

In our recent work \cite{BoGBTr,GBTr}, we followed a reverse route consisting in starting with a fully classical formulation and then quantizing one of the subsystems. The starting point is the Koopman-van Hove (KvH) equation considered in Part I. After writing the KvH equation for e.g. a two-particle wavefunction $\Psi(z_1,z_2)$, we enforce $\partial\Psi/\partial {p_2}=0$ and then apply Dirac's canonical quantization rule $p_2\to - {\rm i}\hbar\partial/\partial q_2$. Here, the notation is such that $z_i=(q^i,p_i)$. In the case of only one particle, one verifies that this process leads to the standard Schr\"odinger equation ${\rm i}\hbar\partial_t\Psi=-({m^{-1}\hbar^2}/{2})\Delta\Psi  +V\Psi$ \cite{Klein}. On the other hand, in the two-particle case one obtains the \emph{quantum-classical wave equation}
\begin{equation}\label{hybrid_KvH}
{\rm i}\hbar\partial_t\Upsilon=\{{\rm i}\hbar \widehat{H},\Upsilon\} - {L}_ {\widehat{H}} \Upsilon, \qquad \text{with} \qquad    {L}_ {\widehat{H}}:=\mathcal{A}\!\cdot\! X_{\widehat{H}} -\widehat{H}
=p\cdot\partial_p\widehat{H}-\widehat{H}
\,.
\end{equation}
Here, $\Upsilon \in \mathscr{H}_{\scriptscriptstyle CQ}=L^2(T^*Q\times M,\mathbb{C})$ is a wavefunction on the hybrid coordinate space $T^*Q\times M$ comprising both the classical and the quantum coordinates $z_1$ and $q_2$, now relabelled simply by $z=(q,p)\in T^*Q$ and $x\in M$, respectively. The function $\widehat{H}(z)$ is defined on $T^*Q$ and takes values in the space of unbounded Hermitian operators on the quantum Hilbert space $\mathscr{H}_{\scriptscriptstyle Q}=L^2( M,\mathbb{C})$. Also, we recall that ${\cal A}=p_i\de q^i\in \Lambda^1(T^*Q)$ is the canonical one form on $T^*Q$.

After reviewing the main aspects of the quantum-classical wave equation, here we shall extend the treatment in Part I by extending  Madelung's hydrodynamic description to hybrid quantum-classical systems. Importantly, this unlocks the door to the  hybrid continuity equation for mixed quantum-classical densities. Eventually, we shall extend the treatment to von Neumann operators and show how the latter may be used to treat more general classes on mixed quantum-classical systems.

\vspace{-0.2cm}

\section{The quantum-classical wave equation}\label{Hybrid_KvH}
This section reviews the algebraic and geometric structure of the quantum-classical wave equation \eqref{hybrid_KvH}. First, we introduce the \textit{hybrid Liouvillian operator}
\begin{equation}\label{QCliouv}
\widehat{\cal L}_{\widehat{H}}=\{{\rm i}\hbar \widehat{H},\ \} - {L}_ {\widehat{H}},
\end{equation}
which is an unbounded self-adjoint operator on $\mathscr{H}_{\scriptscriptstyle CQ}$ thereby making the hybrid wavefunction $\Upsilon(z,x)$ undergo unitary dynamics. The name is due to the fact that in the absence of the hybrid Lagrangian ${L}_ {\widehat{H}}$, this reduces to the classical Liouvillian operator appearing in Koopman's early work. Also, we notice that in the absence of quantum coordinates the hybrid Liouvillian reduces to the prequantum operator from the classical KvH theory.

At this point, it is evident that equation \eqref{hybrid_KvH} is Hamiltonian as it carries the same canonical Poisson structure as the quantum Schr\"odinger equation. In addition, its Hamiltonian  functional (total energy) is conveniently written as
\begin{equation}\label{hybHam}
h(\Upsilon)=\int_{T^*Q}\!\big\langle\Upsilon\big|\widehat{\cal L}_{\widehat{H}}\Upsilon\big\rangle\,{\rm d}z
=
\int_{T^*Q}\!\int_M\big(\bar \Upsilon\,\widehat{\cal L}_{\widehat{H}}\,\Upsilon\big)\,{\rm d}z\wedge {\rm d}x\,.
\end{equation}
Here, the $L^2-$inner product is the immediate extension of the classical and quantum cases. We proceed by emphasizing certain specific aspects of the quantum-classical wave equation. 

\smallskip\noindent
{\bf Properties of the hybrid Liouvillian.} 
As opposed to the prequantum operator from the  KvH theory the hybrid Liouvillian \eqref{QCliouv} does not comprise a Lie algebra structure. Indeed, while prequantum operators satisfy $[{\widehat{\cal L}}_{{H}},{\widehat{\cal L}}_{{F}}]={\rm i}\hbar{\widehat{\cal L}}_{\{H,F\}}$, hybrid Liouvillians possess the following  noncommutative variant:
\begin{equation}\label{noncomm_tuv} 
\big[{\widehat{\cal L}}_{\widehat{H}},{\widehat{\cal L}}_{\widehat{F}}\big]+
\big[{\widehat{\cal L}}_{\bar{H}},{\widehat{\cal L}}_{\bar{F}}\big]^{\!{\sf T}}
=\ {\rm i}\hbar
{\widehat{\cal L}}_{\{{\widehat{H}},{\widehat{F}}\}-\{{\widehat{F}},{\widehat{H}}\}}.
\end{equation}
Here, the bar denotes conjugate operators while $\sf T$ stands for transposition \cite{GBTr}. In addition, the hybrid Liouvillian enjoys relevant equivariance properties with respect to classical and quantum transformations. In particular, if $(\eta, e^{{\rm i} \varphi }) \in \operatorname{Aut}_{ \mathcal{A} }(T^*Q \times S ^1 )$, we have:
\begin{equation}\label{ignazio}
U_{(\eta,e^{{\rm i}\varphi })}^\dagger\widehat{\cal L}_{\widehat{A}}\,U_{(\eta,e^{{\rm i}\varphi })}=\widehat{\cal L}_{\eta^*\!\widehat{A}}
\,,\quad\forall\; (\eta,e^{{\rm i}\theta})\in \widehat{\operatorname{Diff}}_\omega(T^*Q)
\,.
\end{equation}
Here, $\operatorname{Aut}_{ \mathcal{A} }(T^*Q \times S ^1 )$ denotes the group of connection-preserving automorphisms of the prequantum bundle, as explained in Part I.
Alternatively, one also has equivariance under the group ${\mathcal{U}}(\mathscr{H}_{\scriptscriptstyle Q})$ unitary transformations of the quantum Hilbert space space $\mathscr{H}_{\scriptscriptstyle Q}$, namely
\begin{equation}\label{cov2L}
U^\dagger\widehat{\cal L}_{\widehat{A}}U=\widehat{\cal L}_{U^\dagger\widehat{A}U}
\,,\quad\forall\;U\in {\mathcal{U}}(\mathscr{H}_{\scriptscriptstyle Q})
\,.
\end{equation}
These equivariance relations also apply  to  a hybrid density operator extending the classical Liouville density as well as the quantum density matrix.

\smallskip\noindent
{\bf The hybrid density operator.} We define the \textit{hybrid quantum-classical density operator} $\widehat{ \mathcal{D} }$ associated to $\Upsilon$ in such a way that the identity
\[
 \int_{T^*Q}\!\big\langle\Upsilon\big|\widehat{\cal L}_{\widehat{A}}\Upsilon\big\rangle\,{\rm d}z\,=\operatorname{Tr}\int_{T^*Q\!}\widehat{A}\, \widehat{\cal D} \,{\rm d}z,
\]
holds for any hybrid quantum-classical observable $\widehat{A}(z)$. One gets the expression
\begin{equation}\label{hybridDenOp}
\widehat{\cal D}(z)=\Upsilon(z)\Upsilon^\dagger(z) - \operatorname{div}\big( \mathbb{J}{\mathcal{A}} (z)\Upsilon(z) \Upsilon^\dagger(z)\big) + {\rm i}\hbar\{\Upsilon(z),\Upsilon^\dagger(z)\}\,,
\end{equation}
so that $\operatorname{Tr}\int_{\scriptscriptstyle T^*\!Q}\widehat{\cal D}(z)\,{\rm d}z=1$. 
Here, the superscript $\dagger$ denotes the quantum adjoint so that $\Upsilon_1^\dagger(z)\Upsilon_2(z)=\int_M\bar{\Upsilon}_1(z,x)\Upsilon_2(z,x)\,\de x$. Even if unsigned, the operator $ \widehat{ \mathcal{D }}(z)$ represents the hybrid counterpart of the Liouville density in classical mechanics and of the   density matrix in quantum mechanics.
This hybrid operator enjoys the following classical and quantum equivariance properies
\begin{equation}\label{cov1}
\widehat{\cal D}(U_{(\eta,e^{{\rm i} \varphi })}\Upsilon)= \widehat{\cal D}(\Upsilon) \circ \eta ^{-1} 
\,,\quad\forall\; (\eta,e^{{\rm i}\varphi })\in \operatorname{Aut}_ \mathcal{A} (T^*Q \times S ^1 ) 
\,.
\end{equation} 
\begin{equation}\label{cov2}
\widehat{\cal D}(\widehat{{U}}\Upsilon)= \widehat{{U}}\widehat{\cal D}(\Upsilon)\widehat{{U}}^\dagger
\,,\quad\forall\;\widehat{{U}}\in {\mathcal{U}}(\mathscr{H}_{\scriptscriptstyle Q})
\,.
\end{equation} 
The equivariance properties \eqref{cov1}-\eqref{cov2} of the hybrid density operator under both classical and quantum transformations have long been sought in the theory of hybrid classical-quantum systems \cite{boucher} and stand as one of the key geometric properties of the present construction. 

\smallskip\noindent
{\bf Quantum and classical densities.} Given the hybrid density operator, the classical density and the quantum density matrix are obtained as
\begin{equation}
\rho_c(z)=
\operatorname{Tr}\widehat{\cal D}(z)
\,,\qquad\qquad\quad
\hat\rho:=\int_{\scriptscriptstyle T^*\!Q\!}\widehat{\cal D}(z)\,\de z\,,
\end{equation}
respectively.
While $\hat \rho =\int_{T^*Q}\Upsilon\Upsilon^\dagger\,\de z $ is positive by construction,  at present there is no criterion available to establish whether the dynamics of  $\rho  _c$ preserves its positivity, unless one considers the trivial case of absence of coupling, that is $\widehat{H}(z)=H(z)+\widehat{H}$. Possible arguments in favor of Wigner-like negative distributions are found in Feynman's work \cite{Feynman}.
So far, all we know is that the hybrid classical-quantum theory presented here is the only available Hamiltonian theory beyond the mean-field approximation that (1) retains the quantum uncertainty principle and (2) allows the mean-field factorization $\Upsilon(z,x)=\Psi(z)\psi(x)$ as an exact solution in the absence of quantum-classical coupling. However, we recently identified infinite families of hybrid Hamiltonians for which both $\hat{ \rho  }$ and $ \rho  _c$ are positive in time. We will get back to this point later on.

\section{Madelung equations and quantum-classical trajectories \label{sec:HybMad}}

In this section we extend the usual Madelung transformation from quantum mechanics to the more general setting of coupled quantum--classical systems.
Following Madelung's quantum treatment, we shall restrict to consider hybrid Hamiltonians of the type
\begin{equation}\label{QHam}
\widehat{H}(q,p, {x})=-\frac{\hbar^2}{2m}\Delta_x + \frac1{2M}|p|^2 + V(q,x)
\,,
\end{equation}
thereby ignoring the possible presence of magnetic fields. Here $\Delta_x$ and the norm $|p|$ are given with respect to Riemannian metrics on $M$ and $Q$. In this case, the hybrid quantum--classical wave equation \eqref{hybrid_KvH} reads
\begin{equation}
{\rm i}\hbar\partial_t\Upsilon=-\left(L_I+\frac{\hbar^2}{2m}\Delta_x\right)\Upsilon+{\rm i}\hbar\left\{H_I,\Upsilon\right\}
\,,
\label{CQwaveq}
\end{equation}
where we have defined the following scalar functions $L_I, H_I$ on the hybrid space $T^*Q\times M$:
\beq\label{intHam}
H_I(q,p,{x}): = \frac1{2M}|p|^2 + V(q,x)
\,,\qquad\quad
L_I(q,p,{x}): = \frac1{2M}|p|^2 - V(q,x)
\,.
\eeq
These are respectively the classical Hamiltonian and Lagrangian both augmented by the presence of the interaction potential.

\smallskip\noindent
{\bf Madelung transform.} 
We apply the Madelung transform  by writing the hybrid wavefunction in polar form, that is
$\Upsilon = \sqrt{D} e^{{\rm i}S/\hbar}$. Then, the quantum--classical wave equation \eqref{CQwaveq} produces the following hybrid dynamics 
\begin{align}\label{HybHJ}
&\frac{\partial  S}{\partial t}+\frac{|\nabla_{\!x} S|^2}{2m}-\frac{\hbar^2}{2m}\frac{\Delta_x  \sqrt{D}}{\sqrt{D}}= L_I+\left\{H_I, S\right\}\,,
\\
&\frac{\partial  D}{\partial t}+\frac{1}{m}\operatorname{div}_x( D\nabla_{\!x} S)= \left\{H_I,  D\right\},\label{HybHJR}
\end{align}
where the operators $\nabla_{\!x}$, $\operatorname{div}_x$, and $\Delta_x= \operatorname{div}_x\nabla_{\!x}$ are defined in terms of the Riemannian metric on $M$. 
Each equation carries the usual quantum terms  on the left-hand side, while the terms arising from KvH classical dynamics appear  on the right-hand side (see  the corresponding equations in Part I). We observe that \eqref{HybHJ} can be written in Lie derivative form as follows:
\begin{align}\label{HybMad1a}
&\left(\partial_t+\pounds_\mathsf{X}\right) S =\mathscr{L}
\,,\qquad\text{ with }\qquad
\mathsf{X}=\left(X_{H_I},{\nabla_{\!x}S}/m\right)
\,.
\end{align}
Here, $X_{H_I}$ is the $x$-dependent Hamiltonian vector field  on $(T^*Q,\omega)$ associated to $H_I$. Moreover, we have defined the (time-dependent) hybrid  Lagrangian
\[
\mathscr{L}:=L_I + \frac{|\nabla_{\!x} S|^2}{2m} + \frac{\hbar^2}{2m}\frac{\Delta_x  \sqrt{D}}{ \sqrt{D}}\,,
\]
in analogy to the so-called \emph{quantum Lagrangian}  given by the last two terms above.  Then, upon  taking the total differential ${\rm d}$ (on $T^*Q\times M$) of \eqref{HybMad1a} and rewriting \eqref{HybHJR} in terms of the density $D$, we may rewrite \eqref{HybHJ}-\eqref{HybHJR} as follows:
\begin{equation}
\left({\partial_t}+\pounds_\mathsf{X}\right) {\rm d}S =\de\mathscr{L}
\,,
\qquad\qquad\ 
 {\partial_t}D+\operatorname{div}(D \mathsf{X}) =0\,.
\label{HybMad}
\end{equation}
Here, the operator $\operatorname{div}$ denotes the divergence operator induced on $T^*Q\times M$ by the Liouville form on $T^*Q$ and the Riemannian metric on $M$. This form of the hybrid Madelung equations has important geometric consequences which we will present below.

\smallskip\noindent
{\bf Hybrid quantum-classical trajectories.}  
Although the hybrid Madelung equations \eqref{HybHJ}-\eqref{HybHJR} are distinctively different from usual equations of hydrodynamic type, the  equations \eqref{HybMad} still lead to a similar continuum description to that obtained in the quantum case. For example, the second equation in \eqref{HybMad} still yields hybrid  trajectories, which may be defined by considering the density evolution $D(t)=(D_0\circ \Phi(t)^{-1})\sqrt{J_{\Phi(t) ^{-1}} }$, where $\Phi(t)$ is the flow of the vector field $\mathsf{X}$ and $J_{\Phi}$  the Jacobian determinant. Then, this flow is regarded as a Lagrangian  trajectory obeying the equation
\begin{equation}\label{HybTraj}
\dot\Phi(t,z,x)= \mathsf{X}(\Phi(t,z,x))
\,,
\end{equation}
which is the hybrid quantum--classical extension of  quantum  trajectories \cite{Bohm}. In turn, the hybrid  trajectories \eqref{HybTraj} are also useful to express  \eqref{HybHJ} in the form
${\frac{\rm d}{{\rm d}t} (  {\cal S}(t,\Phi(t,z,x))  )} = \mathscr{L}(t,\Phi(t,z,x))$,
which is the hybrid analogue of the KvH phase evolution; see Part I.  Additionally, in the absence of classical degrees of freedom, this picture recovers the quantum Bohmian trajectories since in that case the coordinate $z$ plays no role. Similar arguments apply in the absence of quantum degrees of freedom.

\smallskip\noindent
{\bf The symplectic form.} The first equation in \eqref{HybMad} can be rewritten in such a way to unfold the properties of the flow $\Phi\in \operatorname{Diff}(T^*Q\times M)$ on the quantum-classical coordinate space. Since ${\cal A}\in \Lambda^1(T^*Q)$ and $\Lambda^1(T^*Q)\subset\Lambda^1(T^*Q\times M)$, we denote ${\sf A}=p\cdot \de q\in\Lambda^1(T^*Q\times M)$. Likewise, the differential $\de_x:\Lambda^n(M)\to\Lambda^{n+1}(M)$ on $M$ induces a one-form $\de_x V=\nabla_{\!x}V\cdot\de x\in\Lambda^1(T^*Q\times M)$ on the hybrid coordinate space. Then, with a slight abuse of notation, one can rewrite the first in \eqref{HybMad} as
\[
(\partial_t+\pounds_{\sf X})(\de S-{\sf A})=\de(\mathscr{L}-L_I)-\de_x V
\,.
\]
This equation is crucially important for deducing the quantum-classical dynamics of the Poincar\'e invariant. Indeed, integrating over a loop $\gamma_0$ in $T^*Q\times M$ leads to
\begin{equation}
\frac{\de }{\de t}\oint_{\gamma(t)}p
\cdot\de q = \oint_{\gamma(t)}\! \nabla_{\!x} V\cdot \de x\,,
\,
\end{equation}
where  $\gamma(t)=\Phi(t)\circ\gamma_0$. Then, standard application of Stoke's theorem yields the following relation involving the classical symplectic form $\Omega=-\de{\sf A}$:
\begin{equation}\label{SFevol}
\frac{\de }{\de t}\Phi(t)^*\Omega= -\Phi(t)^* {\rm d}\big({\rm d}_x V\big)\,,\quad\ \text{with}\quad\  {\rm d} ({\rm d}_x V)= \frac{\partial^2 V}{\partial q^{j} \partial x^k}\,{\rm d} q^j\wedge {\rm d}x^k
\,,
\end{equation}
which reduces to the usual conservation  $\Phi(t)^*\Omega=\Omega$  in the absence of coupling.

\section{Joint quantum-classical distributions}
In standard quantum mechanics, the emergence of a probability current gives to the evolution of the quantum density distribution $\rho_q(x)=|\psi(x)|^2$ the structure of a continuity equation. This structure transfers to the hybrid case for which $\partial_t\rho_q=-\operatorname{div}_{x\!}\int_{T^*Q}(D\nabla_{\!x} S/m)\,\de z$, which follows from \eqref{HybHJR}. Here, we will present the analogue of the continuity equation for joint quantum-classical distributions.

\smallskip\noindent
{\bf Hybrid joint distribution.} 
Denoting by $\mathcal{K}_{ \widehat{ \mathcal{D} }}(z;x,y)$ the kernel of the hybrid density operator \eqref{hybridDenOp}, we first introduce the \textit{joint quantum-classical distribution}: 
\begin{align}\label{def_calD1} 
\mathcal{D} (z,x):=&\, \mathcal{K}_{\widehat{ \mathcal{D} } }(z;x,x) 
\\
=&\, |\Upsilon|^2 - \operatorname{div}\!\big( \mathbb{J} {\mathcal{A}} |\Upsilon|^2\big)  + {\rm i} \hbar\{\Upsilon,\bar\Upsilon\}\,.
\label{def_calD2} 
\end{align} 
This represents the joint density for the position of the system in the hybrid space $T^*Q \times M$. Then, the quantum probability density is given by $\rho_q=\int_{T^*Q}\mathcal{D} (z,x)\,\de z$. As shown in \cite{GBTr},  equation \eqref{def_calD2} identifies a momentum map $\Upsilon\mapsto \mathcal{D}(\Upsilon)$ for the  natural action on $ \mathscr{H}_{\scriptscriptstyle CQ}$ of a slight generalization of the group $\operatorname{Aut}_ \mathcal{A} (T^*Q\times S^1)$, which we are now going to introduce. Denote by ${\cal F}(M,G)$ the space of smooth mappings from $M$ to a group $G$; this space is naturally endowed with a group structure inherited by $G$. Then, we let $G=\operatorname{Aut}_ \mathcal{A} (T^*Q\times S^1)$ so that ${\cal F}(M,\operatorname{Aut}_ \mathcal{A} (T^*Q\times S^1))$ possesses a unitary reperesentation on $ \mathscr{H}_{\scriptscriptstyle CQ}$ that is inherited from the van Hove representation  of $\operatorname{Aut}_ \mathcal{A} (T^*Q\times S^1)$ discussed in \makebox{Part I}. We refer the reader to \cite{GBTr} for further details. In particular, the Lie algebra of ${\cal F}(M,\operatorname{Aut}_ \mathcal{A} (T^*Q\times S^1))$ coincides with the Poisson algebra of smooth phase-space functions $A(z;x)$ that are parameterized by the quantum coordinates  $x\in M$, that is ${\cal F}(M,\mathfrak{aut}_ \mathcal{A} (T^*Q\times S^1))\simeq{\cal F}(M,C^\infty (T^*Q))$. Given $A\in{\cal F}(M,C^\infty (T^*Q))$, the corresponding infinitesimal action on $ \mathscr{H}_{\scriptscriptstyle CQ}$ is then $\Upsilon\mapsto -{\rm i}\hbar^{-1}\widehat{\cal L}_{A}\Upsilon$, so that the associated momentum map $\Upsilon \mapsto{\cal D}(\Upsilon)$ is given by \eqref{def_calD2}.
\begin{figure}
\hspace{-1.1cm}{\small 
\begin{xy}
\xymatrix{
& & & & &  &*+[F-:<3pt>]{
\begin{array}{l}
\text{Hybrid density operator}\\
\widehat{ \mathcal{D}}  \in \operatorname{Den}(T^*Q, \operatorname{Her}(\mathscr{H}_{\scriptscriptstyle Q}))
\end{array}
}
\ar[d]|-{\begin{array}{c}\text{\eqref{def_calD1} }\\ \end{array}}&\\
&
*+[F-:<3pt>]{
\begin{array}{c}
\text{Hybrid wavefunctions}\\
\text{$\Upsilon \in \mathscr{H}_{\scriptscriptstyle C Q}$}\\
\end{array}
}
\ar[rrrrr]|{\begin{array}{c}
\text{Momentum map \eqref{def_calD2}  for}\\
\mathcal{F}\big(M,\operatorname{Aut}_{\cal A}(T^*Q \times S ^1 )\big)\\
\end{array}}
\ar[urrrrr]|-{\begin{array}{c}\text{\eqref{hybridDenOp}}\end{array}} & & & & & *+[F-:<3pt>]{
\begin{array}{c}
\text{Joint distribution}\\
\mathcal{D}\in \operatorname{Den}( T^*Q \times M )
\end{array}
}
}
\end{xy}
}
\caption{Relations between hybrid wavefunctions $\Upsilon$,  the hybrid operator $ \widehat{ \mathcal{D} }$, and the joint density $\mathcal{D} $.}
\vspace{-0.5cm}
\end{figure}
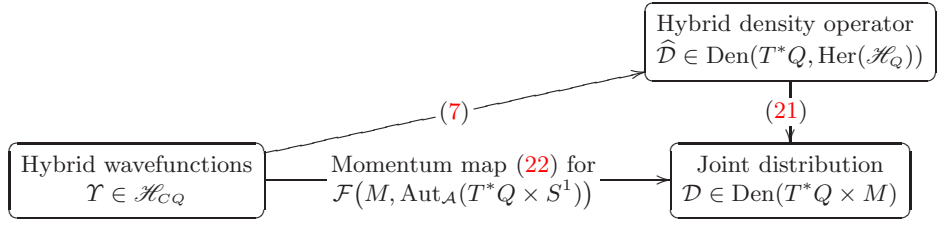

\smallskip\noindent
{\bf Quantum-classical continuity equation.} In analogy to the evolution of the quantum mechanical probability density, the joint distribution satisfies a continuity equation generally involving the hybrid wavefunction. Upon using the polar form $\Upsilon=\sqrt{D}e^{{\rm i}S/\hbar}$, the continuity equation for the joint distribution reads
\begin{equation}\label{CQcont}
\partial_t{\cal D}=-\operatorname{div}{\bf J}=-\operatorname{div}_z J_C-\operatorname{div}_x J_Q
\,,
\end{equation}
where the classical and quantum components of the hybrid current ${\bf J}=(J_C,J_Q)$ are expressed in terms of the vector field ${\sf X}$ in \eqref{HybMad1a} as follows:
\begin{align}
J_C:=&\ {\cal D}X_{H_I}\,,\label{J_C}\\
J_Q:=&\ \frac1m\Big(D\nabla_{\!x} {S}+\partial_{p_i}(p_iD\nabla_{\!x} {S})+\{D\nabla_{\!x}  {S},{S}\}
-\frac{\hbar^2}{4mD}\{D,\nabla_{\!x} D\}
\Big).\label{J_Q}
\end{align}
Notice the emergence of the last term in $J_Q$, which is reminiscent of the quantum potential from standard quantum hydrodynamics. Now, does ${\cal D}$ conserve its initial sign during its evolution? Obviously, this is what happens in the absence of coupling, when $\partial^2_{q^jx^k}V=0$ and the symplectic form $\Omega$ is preserved in \eqref{SFevol}. In addition, if  the quantum kinetic energy is absent in \eqref{QHam}, then $J_Q=0$ and the hybrid continuity equation assumes the characteristic form $\partial_t{\cal D}=\{H_I,{\cal D}\}$ thereby preserving the sign of $\cal D$. Whether this sign preservation extends to more general situations is a subject of current studies. Alternatively, one may ask if the classical density $\rho_c= \int_M\mathcal{D} \,{\rm d}x$ is left positive in time by its equation of motion $\partial_t\rho_c=\int_M\{H_I,{\cal D}\}\,\de x$. The following statement provides a positive answer for certain classes of hybrid Hamiltonians.
\begin{proposition}[\cite{GBTr}]
Assume a hybrid Hamiltonian of the form $\widehat{H}(z)=H(z,\widehat{\alpha})$, where the dependence on the purely quantum observable $\widehat{\alpha}$ is analytic. Assume that the hybrid density operator $\widehat{\cal D}(z)$ is initially positive, then the density $\rho_c$ is also initially positive  and its sign is preserved by the hybrid wave equation \eqref{hybrid_KvH}.
\end{proposition}

\section{A simple closure model}
The high  dimensionality  involved in mixed quantum-classical dynamics poses formidable computational challenges that are typically addressed by devising appropriate closure schemes. The simplest closure is derived by resorting to the mean-field factorization $\Upsilon(z,x)=\Psi(z)\psi(x)$, as described in \cite{BoGBTr}. Alternatively, one can resort to the von Neumann operator description: after defining a quantum-classical von Neumann operator $\widehat{\Xi}$ satisfying ${\rm i}\hbar\partial_t\widehat{\Xi}=[\widehat{\cal L}_{\widehat{H}},\widehat{\Xi}]$, a mean-field factorization is obtained by writing $\widehat{\Xi}=\widehat{\Theta}\hat{\rho}$. Here, $\widehat{\Theta}$  and $\hat{\rho}$ are von Neumann operators on the classical and the quantum Hilbert  spaces $\mathscr{H}_{\scriptscriptstyle C}$ and $\mathscr{H}_{\scriptscriptstyle Q}$, respectively. Then, as discussed in Part I, one may set the integral kernel of $\widehat{\Theta}$ to be ${\cal K}_{\widehat{\Theta}}(z,z')=D(z/2+z'/2)\,e^{\frac{\rm i}{2\hbar}(p+p')\cdot(q-q')}$. 

Here, we will illustrate a slight extension of the mean-field factorization that may again be obtained  by following the final discussion in Part I. First, we perform a convenient abuse of notation by writing $\widehat{\Xi}(z,x,z',x'):={\cal K}_{\widehat{\Xi}}(z,x,z',x')$. Then, the quantity ${\rm i}\hbar\tilde\rho:={\rm i}\hbar \widehat{\Xi}(z,x,z',x')|_{z=z'}$ emerges as  a momentum map for the natural action of $\mathcal{F}(T^*Q,{\cal U}(\mathscr{H}_{\scriptscriptstyle Q}))$ on the Hilbert space $\mathscr{H}_{\scriptscriptstyle CQ}$, while the quantum density matrix is $\int\widehat{\Xi}(z,x,z',x')|_{z=z'}\,\de z$ (again, notice the slight abuse of notation on the integral symbol). Likewise, the hydrodynamic variables
\begin{equation}\nonumber
(\sigma(z),D(z))=\Big(\frac{{\rm i}\hbar}2\frac{\partial}{\partial z}\!\int\!\widehat{\Xi}(z^{\prime\!},x,z,x)\,\de x-\frac{{\rm i}\hbar}2\frac{\partial}{\partial z}\!\int\!\widehat{\Xi}(z,x,z^{\prime\!},x)\,\de x,\,{\widehat{\Xi}}(z,z')\Big)\Big|_{z'=z}
\label{Madmomap2}
\end{equation}
comprise a momentum map structure associated to the action of the semidirect-product group $\operatorname{Diff}(T^*Q) \,\circledS\, \mathcal{F} (T^*Q, S ^1 )$ discussed in Part I. Overall, the map $\widehat\Xi\mapsto(\sigma,D,{\rm i}\hbar\tilde\rho)$ identifies an equivariant momentum map for the action of the semidirect-product $\operatorname{Diff}(T^*Q) \,\circledS\, \big(\mathcal{F} (T^*Q, S ^1 )\times\mathcal{F}(T^*Q,{\cal U}(\mathscr{H}_{\scriptscriptstyle Q})) \big)$. 
At this point,  we express the total energy $h(\widehat{\Xi})=\operatorname{Tr}(\widehat{\Xi}\widehat{\cal L}_{\widehat{H}})$ in terms of the latter momentum map by using the following ansatz (denote $u=\sigma/D$)
\begin{equation}
\widehat{\Xi}(z,x,z',x')=\tilde\rho\big({z/2+z^{\prime\!}/2},x,x'\big)\,e^{\frac{\rm i}{\hbar}(z-z')\cdot u(z/2+z^{\prime\!}/2)}
\,.
\label{ansatz}
\end{equation}
Then, the Hamiltonian functional reads $h(\sigma,D,\tilde\rho)=\operatorname{Tr}\int\!\tilde\rho(D^{-1} X_{\widehat{H}}\cdot\sigma-{L}_{\widehat{H}})\,\de z$ and the resulting system for the dynamics of $(\sigma,D,{\rm i}\hbar\tilde\rho)$ is Lie-Poisson on the dual of the semidirect-product Lie algebra $\mathfrak{X}(T^*Q) \,\circledS\, \big(\mathcal{F} (T^*Q)\times\mathcal{F}(T^*Q,\mathfrak{u}(\mathscr{H}_{\scriptscriptstyle Q})) \big)$. Upon introducing $\hat\rho=\tilde\rho/D$ and $\langle\widehat{A}\rangle=\operatorname{Tr}(\widehat{A}\hat\rho)$, the  equations of motion read
\begin{align*}
&\,\big(\partial_t+\pounds_{\langle X_{\widehat{H}}\rangle}\big)\left(u-{\cal A}\right)
=
(u-{\cal A})\cdot \operatorname{Tr} \!\big(X_{\widehat{H}}\nabla_{\!z} \hat\rho\big)\,,
\\
&\,\partial_tD+\operatorname{div}_{z\!}\big( \langle X_{\widehat{H}}\rangle D\big)=0
\,,\qquad\quad\ 
\partial_t \hat\rho+\langle X_{\widehat{H}}\rangle\cdot\nabla_{\!z} \hat\rho=\big[u\cdot X_{\widehat{H}}-L_{\widehat{H}},\hat\rho\big]
\,,
\end{align*}
where we used $\pounds_{\langle X_{\widehat{H}}\rangle}{\cal A}=\nabla_{\!z}\langle L_{\widehat{H}}\rangle
+\operatorname{Tr}( {\widehat{H}}\nabla_{\!z} \hat\rho)$. Then, setting $u={\cal A}$ yields
\[
\partial_tD+\operatorname{div}_{z}( D\langle X_{\widehat{H}}\rangle)=0
\,,\qquad\qquad
\partial_t \hat\rho+\langle X_{\widehat{H}}\rangle\cdot\nabla_{\!z} \hat\rho=\big[{\widehat{H}},\hat\rho\big]
\,.
\]
Notice that, while $\hat\rho$ must be positive, the operator $\widehat\Xi$ in \eqref{ansatz} is unsigned. Also, the classical density $\rho_c=D$ does not follow a Hamiltonian flow, while the quantum density operator $\hat\rho$ evolves unitarily in the  \makebox{frame moving with  velocity $\langle X_{\widehat{H}}\rangle$.}

\end{document}